\documentclass[preprint,AIP]{revtex4}
\usepackage{amsmath,amssymb}

\def\bi{\boldsymbol{i}}

\begin{document}

\author{Hironobu Kihara\footnote{ hironobu.kihara(at)gmail.com}}

\affiliation{
Research and Education Center for Natural Sciences,
Keio University,\\ 
4-1-1 Hiyoshi, Yokohama,
Kanagawa 223-8521, Japan\\
Faculty of Business and Commerce, Keio University, 
2-15-45 Mita, Minato, Tokyo 108-8345, Japan\\
Faculty of Science and Technology, Seikei University, 
3-3-1 Kichijoji-Kitamachi, Musashino, Tokyo 180-8633, Japan\\
Osaka City University, Advanced Mathematical Institute (OCAMI),
3-3-138 Sugimoto, Sumiyoshi, Osaka 558-8585, Japan\\
Faculty of Science, Ehime University, 
10-13 Dogo-himata, Matsuyama, Ehime 790-8577, Japan
}

\title{Generalized Self-Duality Equations of Polynomial Type in Yang-Mills Theories}

\begin{abstract}
The purpose of this paper is to generalize the self-duality equation by Tchrakian and Corrigan {\it et. al.}. 
Novel generalized self-duality equations on higher-dimensional spaces are discussed. 
This class of equations includes the usual self-duality equation for four-dimensional spaces. 
Some of the generalized self-duality equations over-determine configurations and the existence of solutions is not trivial. 
Several examples of solutions of the equations are demonstrated. 
As an application of the equations, it is proved that some of those solutions solve
the equations of motion derived from rotationally-invariant actions, which consist of single-trace terms and are 
second-order in the time derivative. 
\end{abstract}

\maketitle


\section{Introduction}
Studying solutions of Yang-Mills theory is important in both physics and mathematics. 
The solution by Belavin {\it et. al.} is a milestone in the study of classical solutions of Yang-Mills theory \cite{Belavin:1975fg}. 
That solution has been shown to satisfy self-duality equation. 
Such equations enrich our understanding of the Yang-Mills equation, which is a non-linear partial differential equation. 
Tchrakian considered its generalization to spaces for which the dimensions are greater than four \cite{Tchrakian:1978sf}.  
Grossman, Kephart, and Stasheff also considered similar generalizations in eight-dimensional spaces \cite{Grossman:1984pi}. 
Brihaye, Devchand, and Nuyts also obtained explicit spherically-symmetric solutions of self-duality for eight-dimensional SO(8) gauge theories by using group theoretical techniques \cite{Brihaye:1985mw}. 
We encountered their solution in the framework of string theory and the matrix model \cite{Chen:1999ab}. 
A further generalization of the self-duality equation of this type was discussed by Bais and Batenburg \cite{Bais:1985ns}. 
In \cite{Bellorin:2000dw}, Bellor\'in and Restuccia obtained exact solutions of the Born-Infeld equations by using the generalized self-duality equation. 
By reconsidering Tchrakian's self-duality equation on a six-dimensional sphere, we were able to obtain exact equations for this sphere \cite{Kihara:2007di}. 
There is another generalization of the self-duality equation. 
Corrigan {\it et. al.} considered a linear equation in the field strength \cite{Corrigan:1982th}. 
Their equation has been applied in many scenarios. 

Cremmer and Scherk have already used the configuration solving modified Tchrakian's self-duality equation in the framework of the spontaneous compactification, although they did not consider the self-duality property \cite{Cremmer:1976ir}. 
We applied these to the dynamical compactification and showed that the self-duality equation works to confirm the absence of any tachyonic modes in the gauge sector \cite{Kihara:2009ea,Chingangbam:2009jy}. 

Tchrakian's generalization is a natural extension of the self-duality relation in four-dimensional space. 
He used monomials of field strength 2-form with unit coefficient. 
As we showed in \cite{Kihara:2007di}, the coefficient in general is not unity.  
Because of the nonlinearity in field strength, it is hard to solve and there are few examples of solutions, 
whereas there are many papers following up on the equations by Corrigan {\it et. al.}. 
However their equations seem to require additional geometrical information, for instance K\"ahler structure. 
In this sense, the equations by Corrigan {\it et. al.} can be discussed only on special spaces.
 
There is a simple extension which includes both Tchrakian's equation and equations by Corrigan {\it et. al.}. 
This extension is explained in this paper which is organized as follows. 
In Sec. \ref{sec:gsde} we discuss the generalized self-duality equation developed using polynomials in the field strength. 
In Sec. \ref{sec:sol} we provide several examples of solutions of generalized self-duality equations on some simple spaces. 
In Sec. \ref{sec:summ} we summarize this paper. 
 
\section{Generalized self-duality equations}
\label{sec:gsde}

In this section, we consider Yang-Mills theory on an $n$-dimensional manifold ${\cal M}$ with gauge group $G$. 
The infinitesimal generators of $G$ are denoted by $T_a , (a=1, \cdots , \dim G)$, represented by matrices that satisfy the commutation relations of the Lie algebra of $G$. 
Let $\zeta = (\zeta^1 , \cdots ,\zeta^{n})$ parametrize some open set of the space ${\cal M}$. 

Suppose that the metric is expandable as a sum of symmetric products of vielbeins,  
\begin{align}
ds^2 &=  V^{\mu} \cdot V^{\mu}~,~~(\mu=1,2,\cdots, n)~.
\end{align}

For simplicity in notation, we often use the multiple-index form for vielbeins, 
\begin{align}
V^{\mu_1,\mu_2, \cdots, \mu_p} &{\equiv}
V^{\mu_1} \wedge V^{\mu_2} \wedge \cdots \wedge V^{\mu_p} ~,~~~(\mu_i=1,\cdots, n ; i=1,\cdots, p) 
\end{align}
where $\wedge $ denotes the wedge product, here between vielbeins . 
By using the vielbein, the Hodge dual operator acting on the basis of $p$-forms is defined as 
\begin{align}
* V^{\mu_1 , \cdots , \mu_p}
&= \frac{1}{(n-p)!} \epsilon^{\mu_1\mu_2\cdots \mu_p\nu_{p+1} \cdots \nu_n} 
V^{\nu_{p+1} , \cdots , \nu_n}~. 
\end{align}
where $\epsilon^{\mu_1\cdots \mu_p \nu_{p+1} \cdots \nu_n}$ is the generalized Levi-Civita tensor.

The gauge potential is a one-form $A$ that takes values in the Lie algebra, 
\begin{align}
A_{\mu} &= A_{\mu}^{a} T_a V^{\mu}~,~~(\mu=1,\cdots,n ; a = 1,\cdots , {\rm dim}G)~.
\end{align}
The field strength $F$ is a two-form defined as 
\begin{align}
F &= dA + q A \wedge A~, 
\end{align}
where $q$ is the gauge coupling constant. 
The covariant exterior derivative $D$ acting on $p$-form $\omega$ belonging to an arbitrary unitary representation
\begin{align}
\omega &= \frac{1}{p!} \omega_{\mu_1 , \cdots , \mu_p}
V^{\mu_1 , \cdots , \mu_p}
\end{align}
is defined as 
\begin{align}
D \omega &= d \omega + q {\cal R}(A) \omega~,
\end{align}
where ${\cal R}(A)$ is the representation matrix associated with $A$. 
We call $\omega$ a covariantly-constant $p$-form if $\omega$ does not depend on the gauge field $A$ and satisfies the condition 
$D \omega =0$. 

The $p$-th exterior power of the field strength is denoted as
\begin{align}
F^{\wedge p}  &{\equiv} \overbrace{F \wedge \cdots \wedge F}^p ~. 
\end{align}

Let us next consider the formal sum 
\begin{align}
f(F) &:= \sum_{p=0}^{\infty} \alpha_p F^{\wedge p} ~, ~~f(F)^{\dag}= ~f(F)
\end{align}
where $\alpha_p$ are constant parameters. 
Furthermore, consider the following definition of the pseudo-energy with respect to the function $f$,  
\begin{align}
E_f &:= \int_{{\cal M}} {\rm Tr} f(F) \wedge * f(F)~.
\end{align}
This is the non-Abelian version of the action given in Eq. (33) of \cite{Bellorin:2000dw}. 
The corresponding equation of motion is obtained by the standard variational method
\begin{align}
 \sum_{s,t} \alpha_{s+t+1}  F^{\wedge s}  \wedge  D(* f(F)) \wedge  F^{\wedge t} &=0 ~.
\label{eqn:eom}
\end{align}
If a gauge field $A$ satisfies the following condition, 
\begin{align}
*f(F) &= \sum_{s} C_s \wedge F^{\wedge s} ~,~~~D C_s = 0~, 
\label{eqn:gselfdual}
\end{align}
the configuration solves the equation of motion Eq. (\ref{eqn:eom}). Here $C_s$ are formal sums of differential forms with various ranks. 
Let us call Eq. (\ref{eqn:gselfdual}) the generalized self-duality equations of polynomial type. 
This is a simple generalization of the self-duality equation introduced in \cite{Bais:1985ns}.  
Usually, this equation yields more relations than the number of degrees of freedom for the gauge fields. 
It includes the ordinary self-duality equation in four-dimensional space-time, $F= \pm * F$, which is satisfied by the famous Belavin-Polyakov-Schwartz-Tyupkin instanton solution \cite{Belavin:1975fg}. 
Also, this equation includes Tchrakian's self-duality equations \cite{Tchrakian:1978sf}.

Suppose that $C$ is a constant $0$-form that squares to the unit matrix, $C C = {\bf 1}$. 
By taking the Bogomol'nyi completion of the pseudo-energy $E_f$,  
\begin{align}
E_f &= \int_{{\cal M}} {\rm Tr} f(F) \wedge * f(F) \cr
&=   \frac{1}{2}   \int_{{\cal M}} {\rm Tr} \left\{f(F) \wedge * f(F) + *  C f(F) \wedge ** C f(F) \right\} \cr
&= \frac{1}{2} \int_{{\cal M}} {\rm Tr} \left\{ f(F) -  * {\varepsilon} C f(F) \right\}
\wedge *\left\{ f(F) -  * {\varepsilon} C f(F) \right\} 
+ \varepsilon \int_{{\cal M}}  {\rm Tr} C f(F)^{\wedge 2} ~, 
\end{align}
and comparing the second and third expressions, we obtain the corresponding Bogomol'nyi equation, 
\begin{align}
*f(F) &= {\varepsilon} C f(F) ~, 
\label{eqn:gselfdual2}
\end{align}
where $\varepsilon$ is a sign.
This relation is the kind of generalized self-duality equation of polynomial type 
that is alluded to in naming Eq. (\ref{eqn:gselfdual}). 
As a solution of Eq. (\ref{eqn:gselfdual2}), we have 
\begin{align}
E_f &= \varepsilon \int_{{\cal M}}  {\rm Tr} C f(F)^{\wedge 2}~,
\end{align}
which is a topological quantity. 
Let us consider the fluctuation $U$ around a classical solution $A^{(0)}$, $A \rightarrow A' = A^{(0)} + s U$, where $s$ is an auxiliary parameter. 
The covariant derivative of the fluctuation $U$ is written as
$D^{(0)} U  = d U + q ( A^{(0)} \wedge U + U \wedge A^{(0)}  )$. 
The field strength $F'$ is written as $F' = F^{(0)} + s D^{(0)} U + qs^2 U \wedge U$ which induces in the function $f(F')$ a fluctuation given by 
the following formal expansion with respect to the parameter $s$ 
\begin{align}
f(F') &= \sum_{i=0}^{\infty} c_i(F^{(0)},D^{(0)} U ,U) s^i ~. 
\end{align}
where the terms $c_i$ depend on the classical fields. For instance, $c_0 = f(F^{(0)})$ and 
\begin{align}
c_1 &= \sum_{p=1}^{\infty} \alpha_p  \sum_{j=0}^{p-1} F^{(0) , \wedge j} \wedge 
D^{(0)} U \wedge  F^{(0), \wedge p-j-1} ~.
\end{align}
By expanding the pseudo-energy, 
\begin{align}
\delta E & = 2 s \int_{{\cal M}} {\rm Tr} c_1 \wedge * f(F) 
+ s^2 \int_{{\cal M}}  {\rm Tr} \left\{ c_1 \wedge * c_1 + 2 c_2 \wedge * f(F) \right\}   
+ \cdots 
\end{align}
we obtain the effective pseudo-energy in the classical sense. Here, the first term vanishes because the equation of motion enables this to be written as a total derivative. 
The term ${\rm Tr} c_1 \wedge * c_1$ yields the kinetic terms and mass terms of $U$, and the term ${\rm Tr}c_2 \wedge * f(F)$ provides a correction to the mass terms. 
This latter term may lower the mass eigenvalues and tachyonic mass terms may appear. 

Suppose that the configuration $A=A^{(0)}$ solves equation (\ref{eqn:gselfdual2}). 
Let us consider the following combination, 
\begin{align}
{\cal B} &=  f(F) -  * \varepsilon C f(F) ~.
\end{align}
The expansion of ${\cal B}$ with respect to parameter $s$ is 
\begin{align}
\delta {\cal B} &= \sum_{i=0}^{\infty} \left\{ c_i(F^{(0)},D^{(0)} U ,U) - 
* C c_i(F^{(0)},D^{(0)} U ,U)  \right\} s^i ~. 
\end{align}
Because $A^{(0)}$ solves $f(F)= * C f(F)$, the zero-th order term 
$c_0(F^{(0)},D^{(0)} U ,U) - * C c_0(F^{(0)},D^{(0)} U ,U)  $ vanishes. 
Hence 
\begin{align}
\delta {\cal B} &= \sum_{i=1}^{\infty} \left\{ c_i(F^{(0)},D^{(0)} U ,U) - 
* C c_i(F^{(0)},D^{(0)} U ,U)  \right\} s^i ~. 
\end{align}
As this series is greater than zeroth-order in $U$, 
the variation of the pseudo-energy $\delta E$ becomes positive definite,
\begin{align}
\delta E&= \int_{{\cal M}} {\rm Tr}
\delta {\cal B} \wedge *  \delta {\cal B}~. 
\end{align}
This ensures the stability of the configuration $A = A^{(0)}$. 
If $\delta {\cal B}(u) = 0$ for some fluctuation mode $U=u \neq 0$, then we call $u$ a flat direction. 

Let us consider the following pseudo-energy, 
\begin{align}
E_{f,g} &:= \int_{{\cal M}} {\rm Tr} \left\{ f(F) \wedge * f(F) + g(F) \wedge * g(F) \right\} ~.
\end{align}
Suppose that $C$ is covariantly constant with $C^2 = {\bf 1}$, then 
\begin{align}
E_{f,g} &:= \int_{{\cal M}} {\rm Tr} \left\{ f(F)  - * C g(F) \right\} \wedge * \left\{  f(F) -  * C g(F) \right\} 
+ \int_{{\cal M}} {\rm Tr} C f(F) \wedge g(F)~. 
\end{align}
If a configuration solves the equation $f(F) = * C g(F)$, it also solves the equation of motion and is stable as well. In this case, by replacing ${\cal B} =f(F)  - * C g(F) $, the same story works as the above consideration. 
We refer to (\ref{eqn:gselfdual2}) and $f(F) = * C g(F)$ as the Bogomol'nyi equations in the rest of the paper. 
In the next section we provide several examples of the solution of these equations. 

\section{Solutions}
\label{sec:sol}

\subsection{${\mathbb R}^{2m}$}
We begin by considering the generalized self-duality equation on a $2m$-dimensional real vector space. 
The gauge group is U(1). 
\begin{align}
ds^2 &= | d \zeta |^2~,&
V^i &= d \zeta^i
\end{align}

The simplest example is 
\begin{align}
A &= \frac{1}{2} H \sum_{s=1}^u  \left( \zeta^{2s-1} V^{2s} - \zeta^{2s} V^{2s-1} \right)~,\cr
F &= dA = H ( V^{12} + V^{34} + \cdots V^{2u-1,2u} )
\label{eqn:selfdualflat}
\end{align}
where $H$ is a constant. 
We obtain $F^{\wedge p}=0$ ($p>u$). 
For $p\leq u$, 
\begin{align}
F^{\wedge p} &= p! H^p \sum_{a_i}  V^{2a_1-1,2a_1,\cdots 2a_p-1,2a_p}
\end{align}
Hence we have 
\begin{align}
*F^{\wedge p} &=  \frac{p!}{(u-p)!} H^{2p-u}F^{\wedge (u-p)} \wedge 
V^{2u+1,2u+2 , \cdots , 2m-1,2m} 
\end{align}
Because $d V^{\mu}=0$ and $V^{\mu}$ does not carry charges, the coefficient is covariantly constant, 
\begin{align}
*f(F) &= \sum_{p=0}^{u} \alpha_p \frac{p!}{(u-p)!} H^{2p-u}F^{\wedge (u-p)} \wedge 
V^{2u+1,2u+2 , \cdots , 2m-1,2m}   ~.
\end{align}

Let us next consider when this configuration solves Eq. (\ref{eqn:gselfdual2}). 
Because the coefficient should be a 0-form, we require that $u=m$ 
\begin{align}
*f(F) &= \sum_{p=0}^{m} \alpha_p \frac{p!}{(m-p)!} H^{2p-m}F^{\wedge (m-p)}.
\end{align}
Eq. (\ref{eqn:gselfdual2}) also requires that the function $f$ is a particular function that satisfies
\begin{align}
*f(F) &= C f(F)\cr
\alpha_p \frac{p!}{(m-p)!} H^{2p-m} &= C \alpha_{m-p} \cr
\alpha_p p! H^p &= C \alpha_{m-p} (m-p)! H^{m-p}  \cr
\alpha_p &= \frac{L}{p!H^p} ~,~~~C=1~. 
\end{align}
where $L$ is constant. 
Hence we find that the function should be 
\begin{align}
f(F) &= L \sum_{p=0}^{\infty} \frac{1}{p!} \left( \frac{F}{H} \right)^{\wedge p} 
= L \exp\left( \frac{F}{H} \right)  ~. 
\end{align}
The Bogomol'nyi equation of $E_f$ using this function $f$ is $f(F) = *f(F)$ and the gauge configuration
(\ref{eqn:selfdualflat}) solves the equation. 

\subsection{$S^{2m}$}

In this subsection, we consider an SO($2m$) gauge theory on a $2m$-dimensional manifold $S^{2m}$. 
The solution given here has been considered in various contexts, for instance in \cite{Bais:1985ns}. 
The metric and vielbeins are expressed as: 
\begin{align}
ds^2 &=  \frac{ | d \zeta |^2}{(1+ |\zeta |^2/4R^2 )^2}~,\cr
V^i & {\equiv} \frac{d \zeta^i}{(1+ |\zeta |^2/4R^2 )}~,
\end{align}
where $|\zeta|^2 = (\zeta^1)^2 + \cdots + (\zeta^{2m})^2$ and $R$ is  the radius.

We introduce the Clifford algebra of matrices $\gamma_a$ satisfying the anticommutation relations $\{ \gamma_a, \gamma_b\}=2 \delta_{ab}$, where $(a,b=1,2,\cdots ,2m)$.  
The space of infinitesimal generators of SO($2m$) is spanned by the commutators
$\gamma_{a,b} = (1/2) [\gamma_a, \gamma_b]$.
Let us define the chiral matrix $\gamma_{2m+1}$, 
\begin{align}
\gamma_{2m+1} & {\equiv} i^{-m} \gamma_1 \gamma_2 \cdots \gamma_{2m}~,
\end{align}
which satisfies $\gamma_{2m+1}^2=1$.  

We will sometimes use the following notation, 
\begin{align}
\gamma_{a(1),a(2), \cdots, a(p)} &{\equiv}
\frac{1}{p!} \sum_{\sigma \in {\mathfrak S}_p} {\rm sgn}(\sigma)
\gamma_{a(\sigma(1))}  \gamma_{a(\sigma(2))}  \cdots \gamma_{a(\sigma(p))} ~,
~~~(a(i)=1,\cdots, 2m , i=1,\cdots, p) ~.
\end{align}
The gauge potential one-form $A$ takes values in the algebra so($2m$) of the gauge group, 
\begin{align}
A &= \frac{1}{2} A_{\mu}^{a,b} V^{\mu} \gamma_{a,b}~,~~(\mu=1,\cdots,2m , a,b = 1,\cdots ,2m)~.
\end{align}

The covariant exterior derivative $D$ acting on the 
following $p$-form $\omega$ 
\begin{align}
\omega &= \frac{1}{p!}\frac{1}{s!} \omega_{\mu_1 , \cdots , \mu_p}^{a_1, \cdots , a_s} 
V^{\mu_1 , \cdots , \mu_p} \gamma_{a_1, \cdots , a_s}
\end{align}
is defined as 
\begin{align}
D \omega &= d \omega + q ( A \wedge \omega - (-1)^p \omega \wedge A )~.
\end{align}

Let us consider the following gauge potential, 
\begin{align}
A &= \frac{1}{4qR^2} \zeta^a V^b \gamma_{a,b}~.
\label{eqn:selfdualsphere}
\end{align}
The field strength is
\begin{align}
F &= \frac{1}{4qR^2} V^{ab} \gamma_{a,b} 
\end{align}
which yields for the $p$-th power of the field strength 
\begin{align}
F^{\wedge p} &=\frac{1}{(4qR^2)^p} 
V^{a(1),b(1),\cdots,a(p),b(p)} \gamma_{a(1),b(1),\cdots,a(p),b(p)} ~.
\end{align}
This configuration satisfies the following self-dual property, 
\begin{align}
* F^{\wedge p} &=(4qR^2)^{m-2p}\frac{i^m (-1)^{m-p}(2p)!}{(2m-2p)!} \gamma_{2m+1} F^{\wedge (m-p)}
\end{align}
Hence for an arbitrary function $f(F)$, the configuration solves the generalized self-duality equation 
\begin{align}
* f(F) &= \sum_{p=0}^{m} \alpha_p * F^{\wedge p} \cr
&=  \sum_{p=0}^{m} \alpha_p (4qR^2)^{m-2p}\frac{i^m (-1)^{m-p}(2p)!}{(2m-2p)!} \gamma_{2m+1} F^{\wedge (m-p)}~,
\end{align}
because $\gamma_{2m+1}$ commutes with an arbitrary product of generators, $[\gamma_{2m+1} , \gamma_{a,b}] =0$. 

To elaborate,
Eq. (\ref{eqn:gselfdual2}) requires that the function $f$ is the special function satisfying $* f(F) = C f(F) $ which on comparing the coefficients of $F^{\wedge p}$ leads to constraint\begin{align}
\alpha_p (4qR^2)^{m-2p}\frac{i^m (-1)^{m-p}(2p)!}{(2m-2p)!} \gamma_{2m+1} 
&= C \alpha_{m-p} \cr
\frac{(2p)! i^p \alpha_p}{(4qR^2)^p} \gamma_{2m+1} &= C   \frac{(2(m-p))! i^{m-p} \alpha_{m-p}}{(4qR^2)^{m-p}}~,
\end{align}
where the latter expression is a rearrangement of the former. The solutions for $\alpha_p$ and $C$ are obtained as follows, 
\begin{align}
\alpha_p&= L \frac{(4qR^2)^p}{(2p)! i^p} ~,~~~C= \gamma_{2m+1}. 
\end{align}
Hence we obtain
\begin{align}
f(F) &= L \sum_{p=0}^{\infty} \frac{1}{(2p)!} ( - 4qR^2  i F )^{\wedge p} = L \cosh \sqrt{-4qR^2i F}~.
\end{align}
The Bogomol'nyi equation of $E_f$ using this function $f$ is $f(F) = * \gamma_{2m+1}f(F)$ and the gauge configuration (\ref{eqn:selfdualsphere}) solves the equation.

\subsection{$S^2 \times S^2 \times S^2$}
In this subsection, we discuss the generalized self-duality equation on $S^2 \times S^2 \times S^2$ in a U(1) gauge theory. We treat the gauge fields here as real vector fields. 
Let us consider the following metric and vielbeins, 
\begin{align}
ds^2 &=  R_1^2 \frac{ | d x |^2}{(1+ |x |^2/4 )^2}
+  R_2^2 \frac{ | d y |^2}{(1+ |y |^2/4 )^2}
+ R_3^2 \frac{ | d z |^2}{(1+ |z |^2/4 )^2}
~,\cr
V^i &{\equiv} R_1\frac{d x^i}{(1+ |x|^2/4 )}~,\cr
V^{i+2} & {\equiv} R_2\frac{d y^i}{(1+ |y|^2/4 )}~,\cr
V^{i+4} &{\equiv} R_3\frac{d z^i}{(1+ |z|^2/4 )}~,
\end{align}
where $i=1,2$,
and the following gauge configuration, 
\begin{align}
A &=  A^{[1]} + A^{[2]}+A^{[3]}~,\cr
  A^{[1]} &=  C_1 ( x^1 V^2  - x^2 V^1 )~,\cr
 A^{[2]} &=  C_2 ( y^1 V^4  - y^2 V^3 )~,\cr
 A^{[3]} &=  C_3 ( z^1 V^6  - z^2 V^5 )~.
\label{eqn:selfduals2s2s2}
\end{align}
Here $C_{\alpha},~(\alpha=1,2,3)$ are constant parameters. 
The corresponding field strength is 
\begin{align}
F &= F^{[1]} + F^{[2]} + F^{[3]} \cr
F^{[\alpha]} &= d A^{[\alpha]} = \frac{2 C_{\alpha}}{R_{\alpha}} V^{2\alpha-1}\wedge V^{2\alpha} ~,~~~\alpha=1,2,3~~,
\end{align}
for which the square of the field strength expands as 
\begin{align}
F \wedge F &=2 \left(  F^{[2]} \wedge F^{[3]} + F^{[3]} \wedge F^{[1]} + F^{[1]} \wedge F^{[2]}    \right) \cr
&= \frac{8 C_1 C_2 C_3}{R_1R_2R_3}\left(  \frac{R_1}{C_1} V^{3456} +   \frac{R_2}{C_2} V^{1256}  
+  \frac{R_3}{C_3} V^{1234}   \right)
\end{align}
with Hodge dual  
\begin{align}
* F \wedge F &= \frac{8 C_1 C_2 C_3}{R_1R_2R_3}\left(  \frac{R_1}{C_1} V^{12} +   \frac{R_2}{C_2} V^{34}  
+  \frac{R_3}{C_3} V^{56}   \right)~.
\end{align}
The relation $*F \wedge F \propto F$ requires  
\begin{align}
C_{\alpha} &= \frac{R_{\alpha}}{\sqrt{2} \lambda}~,
\label{eqn:coeffs2s2s2}
\end{align}
where $\lambda$ is an arbitrary parameter. Hence we obtain the following self-dual relation, 
\begin{align}
F &= * \frac{\lambda}{2 \sqrt{2}} F \wedge F~.
\end{align}
One can now derive the Bogomol'nyi equation which is obtained from the following pseudo energy, 
\begin{align}
E &= \frac{1}{4} \int_{\cal M}  \left\{  F \wedge * F 
+ \alpha^2 ( F \wedge F ) \wedge * ( F \wedge F )  \right\} ~. 
\label{eqn:penergy-s2s2s2}
\end{align}
This works out to be 
\begin{align}
F &= * \alpha F \wedge F~.
\label{eqn:bogomols2s2s2}
\end{align}
The configuration 
\begin{align}
A &=   \frac{R_{1}}{4\alpha} ( x^1 V^2  - x^2 V^1 )
+ \frac{R_{2}}{4 \alpha} ( y^1 V^4  - y^2 V^3 )
+  \frac{R_{3}}{4 \alpha} ( z^1 V^6  - z^2 V^5 )~
\end{align}
solves Eq. (\ref{eqn:bogomols2s2s2}). 
In this expression we do not have any restriction on the coupling constant $q,\alpha$ and radii $R_{\alpha}$. 

\subsection{$S^4 \times S^2$}

In this final subsection, 
we obtain a new solution for the generalized self-duality equations corresponding to the direct product of a four-dimensional sphere and a two-dimensional sphere, in the sense of Tchrakian. The solution is actually a combination of the Dirac monopole and Belavin-Polyakov-Schwartz-Tyupkin instanton.  
The proportional constant of the self-duality equation is different from that obtained for the six-dimensional sphere. 

We first develop the Bogomol'nyi equations obtained for the U(2) Yang-Mills theory with the following pseudo-energy on a six-dimensional space, denoted here by ${\cal M}$, 
\begin{align}
E &= \frac{1}{4} \int_{\cal M} {\rm Tr} \left\{ - F \wedge * F 
+ \alpha^2 ( F \wedge F ) \wedge * ( F \wedge F )  \right\} ~. 
\label{eqn:penergys2s4}
\end{align}
where now $F$ is the gauge field strength 2-form taking values in the Lie algebra u(2) and $\alpha$ is the coupling constant. 
The Hodge dual operator, $*$, is defined by the metric $ds^2$ on the space ${\cal M}$. 
The vielbein of $ds^2$ is denoted by $V^M$ $(M=1,2,\cdots,6)$.
The Bogomol'nyi equation obtained from the pseudo-energy Eq.(\ref{eqn:penergys2s4}) is 
\begin{align}
F &= \pm * \bi \alpha F \wedge F~.
\label{eqn:self-dual} 
\end{align}
Because $F$ is a 2-form, the coefficients of $F \wedge F$ is written by anti-commutator, 
\begin{align}
F \wedge F &= \frac{1}{8} \{ F_{MN} , F_{PQ} \} V^{MNPQ}~.
\end{align} 
The anti-commutator of two anti-Hermitian matrices is a Hermitian matrix.

The space of $2 \times 2$ anti-Hermitian matrices is spanned by 
\begin{align}
\tau_1 &:=- \begin{pmatrix} 0 & \bi \\ \bi & 0 \end{pmatrix}~,&
\tau_2 &:= -\begin{pmatrix} 0 & 1 \\ -1 & 0 \end{pmatrix}~,&
\tau_3 &:=- \begin{pmatrix} \bi & 0 \\ 0 & - \bi \end{pmatrix}~,&
\tau_4 &:= \begin{pmatrix} \bi & 0 \\ 0 & \bi \end{pmatrix}~. 
\end{align}
The gauge field 1-form $A$ can then be written as 
$A = A_{M}^{\alpha} V^M \tau_{\alpha} $, where $M=1,2,\cdots, 6$ and $\alpha=1,2,3,4$. 
Suppose that ${\cal M}=S^4 \times S^2$ with metric given by
\begin{align}
ds^2 &= R_1^2 \frac{|dx|^2}{(1+|x|^2/4)^2}
+ R_2^2 \frac{|dy|^2}{(1+|y|^2/4)^2} ~.
\end{align}
Here the coordinate system on $S^4$ is denoted by $x=(x^1,x^2,x^3,x^4)$ and the coordinate system on $S^2$ by $y=(y^5,y^6)$. 
\begin{align}
V^a &= R_1 \frac{dx^a}{(1+|x|^2/4)}\cr
V^{i+4} &= R_2 \frac{dy^{i+4}}{(1+|y|^2/4)}
\end{align}
where $a=1,2,3,4$ and $i=1,2$. 
We use notations $|x|^2 := (x^1)^2+ (x^2)^2+ (x^3)^2+ (x^4)^2$ and 
$|y|^2 := (y^5)^2+ (y^6)^2$. 
$R_{1,2}$ are radii of $S^4$ and $S^2$, respectively.  
We use indices, $a,b$, for labeling the directions along $S^4$; 
 indices, $i,j$, label those along $S^2$. 
We use $M,N$, $i,j$ and $a,b$ to index vielbeins.

We introduce the ansatz
\begin{align}
A &=  A^{[1]} + A^{[2]}\cr
  A^{[1]} &= \frac{1}{q} \frac{|x|^2}{|x|^2 + \rho^2} U^{\dag} d U  \cr
  A^{[2]} &=  C ( y^5 V^6  - y^6 V^5 ) \tau_4
\end{align}
where $\rho$ and $C$ are suitable constants and $U$ is a $2 \times 2$ unitary matrix defined as
\begin{align}
U &= \frac{1}{|x|} X~,\cr
X &= x^4 {\bf 1}_2 + x^1 \tau_1 + x^2 \tau_2 + x^3 \tau_3
 ~.
\end{align}

$A^{[1]}$ is the gauge field of the Belavin-Polyakov-Schwartz-Tyupkin instanton and 
$A^{[2]}$ is the gauge field which is the restriction of the Dirac monopole onto $S^2$. 
The field strength $F$ with respect to the gauge field $A$ is 
\begin{align}
F &= dA + q A \wedge A= F^{[1]} + F^{[2]} \cr
F^{[1]}  &= d A^{[1]} + q  A^{[1]}  \wedge  A^{[1]} \cr
&=\frac{\rho^2}{q R_1^2} \frac{(1+|x|^2/4)^2}{(|x|^2 + \rho^2)^2}
 (2V^4 \wedge V^a - \epsilon^{abc4} V^b \wedge V^c ) \tau_a  \cr
F^{[2]} &=   d  A^{[2]}= \frac{2C}{R_2} V^5 \wedge V^6 \tau_4
\end{align}
The square of $F^{[1]}$ can be expressed as
\begin{align}
F^{[1]} \wedge F^{[1]} &= - \left\{\frac{\rho^2}{q R_1^2} \frac{(1+|x|^2/4)^2}{(|x|^2 + \rho^2)^2} \right\}^2
 (2V^4 \wedge V^a - \epsilon^{abc4} V^b \wedge V^c ) \wedge (2V^4 \wedge V^{a} - \epsilon^{ab'c'4} V^{b'} \wedge V^{c'} ) {\bf 1}_2\cr
&= + \left\{\frac{\rho^2}{q R_1^2} \frac{(1+|x|^2/4)^2}{(|x|^2 + \rho^2)^2} \right\}^2
4V^4 \wedge V^a \wedge \epsilon^{abc4} V^{b} \wedge V^{c} {\bf 1}_2 \cr
&= - 4!  \left\{\frac{\rho^2}{q R_1^2} \frac{(1+|x|^2/4)^2}{(|x|^2 + \rho^2)^2} \right\}^2 V^1 \wedge V^2 \wedge V^3 \wedge V^4
{\bf 1}_2
~.
\end{align}
As $F^{[1]}$ and $F^{[2]}$ are two-forms and $F^{[2]}$ is proportional to unit matrix ${\bf 1}_2$, 
$F^{[1]}$ and  $F^{[2]}$ commute with each other. 
Hence we obtain
\begin{align}
F \wedge F &= F^{[1]} \wedge F^{[1]}  + 2 F^{[1]} \wedge F^{[2]}~.
\end{align}
Let us consider the self-duality equation. 
The Hodge dual of $F \wedge F$ is separate into two parts,
\begin{align}
* F \wedge F &= * F^{[1]} \wedge F^{[1]}  + * 2 F^{[1]} \wedge F^{[2]}
\end{align}
The separate terms are written as
\begin{align}
* F^{[1]} \wedge F^{[1]} &=   4! \bi  \left\{\frac{\rho^2}{q R_1^2} \frac{(1+|x|^2/4)^2}{(|x|^2 + \rho^2)^2} \right\}^2 
\frac{R_2}{2C} F^{[2]} \cr
* 2 F^{[1]} \wedge F^{[2]} &= \bi \frac{4C}{R_2} \frac{\rho^2}{q R_1^2} \frac{(1+|x|^2/4)^2}{(|x|^2 + \rho^2)^2}
 (\epsilon^{4abc} V^b \wedge V^c  - 2 V^a \wedge V^4 ) \tau_a \cr
&= \frac{4C}{R_2} \bi F^{[1]} ~, 
\end{align}
And hence we obtain 
\begin{align}
- * \bi F \wedge F  &=  4!  \left\{\frac{\rho^2}{q R_1^2} \frac{(1+|x|^2/4)^2}{(|x|^2 
+ \rho^2)^2} \right\}^2 \frac{R_2}{2C} F^{[2]}+  \frac{4C}{R_2} F^{[1]}  ~.
\end{align}
Anticipating the equation (\ref{eqn:self-dual}), the coefficients have to be constant which implies that $\rho =2$. The above equation (71) reduces to
\begin{align}
- * \bi F \wedge F  &= 4!  \left\{\frac{1}{4q R_1^2}  \right\}^2 \frac{R_2}{2C} F^{[2]}
+  \frac{4C}{R_2} F^{[1]} ~. \cr
&= \frac{3R_2}{4q^2CR_1^4}F^{[2]}
+  \frac{4C}{R_2} F^{[1]} \cr
&= \frac{4C}{R_2} \left( F^{[1]} + \frac{3R_2^2}{16q^2C^2 R_1^4}F^{[2]} \right) 
\end{align}
The last expression should be proportional to $F$. Hence the constant $C$ is determined as
\begin{align}
C &= \varepsilon \frac{\sqrt{3}R_2}{4q R_1^2} ~,
\end{align}
where $\varepsilon=+1$ or $-1$. 
The gauge field satisfies the following self-duality equation, 
\begin{align}
 *  F \wedge F  &= \varepsilon \bi \frac{\sqrt{3}}{q R_1^2} F ~,
\end{align}
where we note that the proportional constant does not depend on the radius $R_2$.
The gauge field is given as, 
\begin{align}
A &=  A^{[1]} + A^{[2]}\cr
  A^{[1]} &= \frac{1}{q} \frac{|x|^2}{|x|^2 + 4} U^{\dag} d U  \cr
  A^{[2]} &=  \varepsilon \frac{\sqrt{3}R_2}{4q R_1^2} ( y^5 V^6  - y^6 V^5 ) \tau_4 ~.
\end{align}

If the coupling constant takes the specific value, $\displaystyle \alpha = \frac{q R_1^2}{\sqrt{3}}$, the self-duality equation becomes the Bogomol'nyi equation,  Eq. (\ref{eqn:self-dual}). 
It is interesting to compare this with the value $\displaystyle \alpha = \frac{q R_2^2}{3}$ appearing in \cite{Kihara:2007di}. 
The constant 
$J$ ($F=*J F \wedge F$)
depends on the underlying space, gauge coupling constant, and the gauge configuration. 
If one requires that the configuration satisfies the Bogomol'nyi equation, the radii of $S^6$ and $S^4$ are determined by coupling constants $\alpha$ and $q$, while the radius of $S^2$ remains a free parameter of the solutions. 

\section{Summary and Discussion}
\label{sec:summ}
The generalized self-duality equation of polynomial type was discussed. 
The generalized self-duality equation helps to solve the equations of motion of Yang-Mills theories. 
The Bogomol'nyi equation derived from the pseudo-energy defined by either one polynomial or two polynomials was regarded as a generalized self-duality equation. 
Within the solutions of the Bogomol'nyi equation, there are no tachyonic modes. 
Examples corresponding to spaces ${\mathbb R}^{2m},S^{2m}$, $S^2 \times S^2 \times S^2$ and $S^2 \times S^4$ were presented. 
Requiring for the examples that the solutions of self-duality equations are also solutions of Bogomol'nyi equations determines the form of the pseudo-energy. 
We also observed that the proportionality constants of these self-duality equations depend on geometrical parameters of the spaces.  

Further extension of models including non-single trace terms is required, as shown in the case of complex projective spaces \cite{Kihara:2008zg}.   
The application of the technique to various models involving the decomposition of the representation as discussed in \cite{Brihaye:1985mw} is very attractive.  
The physical application of the solutions on $S^2 \times S^4$ to dynamical compactification may be interesting. 
There a discussion on the violation of isotropy not only for the compact directions but also over the total space including noncompact directions is needed in such an application.

\begin{acknowledgments}
I am grateful to Jorge Alejandro Bellor\'in Romero and Chandrashekar Devchand for their comments. 
The professional editing service of Edanz Group Japan has much improved this article.  
This work is supported by Hujukai, Iwanami-shoten and Seikei University.  
\end{acknowledgments}

\end{document}